\documentclass[iop]{emulateapj}

\usepackage{txfonts}
\usepackage{graphicx}
\usepackage{graphicx}
\usepackage{natbib}

\usepackage{natbib}

\shorttitle{Characteristics of transverse waves in chromospheric mottles}
\shortauthors{Kuridze et al.}

\begin{document}

\title{Characteristics of transverse waves in chromospheric mottles}
\vskip1.0truecm
\author{
D. Kuridze$^{1,5}$, G. Verth$^{2}$, M. Mathioudakis$^{1}$, R. Erd\'elyi$^{2}$, D. B. Jess${^1}$, R. J. Morton$^{3}$, D. J. Christian$^{4}$, F. P. Keenan${^1}$}
\affil{
$^1$Astrophysics Research Centre, School of Mathematics and Physics, Queen's University Belfast, BT7~1NN, Northern Ireland, UK; e-mail: dkuridze01@qub.ac.uk} 
\affil{$^2$Solar Physics and Space Plasma Research Centre (SP$^2$RC), University of Sheffield, Hicks Building, Hounsfield Road, Sheffield S3 7RH, UK}
\affil{$^3$Mathematics and Information Science, Northumbria University, Camden Street, Newcastle Upon Tyne, NE1 8ST}
\affil{$^4$Department of Physics and Astronomy, California State University, Northridge, CA 91330, USA}
\affil{$^5$Abastumani Astrophysical Observatory at Ilia State University, G. Tsereteli 3, 0612, Tbilisi, Georgia}

\begin{abstract}
Using data obtained by the high temporal and spatial resolution Rapid Oscillations in the Solar Atmosphere (ROSA) instrument on the 
Dunn Solar Telescope, we investigate at an unprecedented level of detail transverse oscillations in chromospheric  fine structures near the solar disk center. 
The oscillations are interpreted in terms of propagating and standing magnetohydrodynamic kink waves. Wave characteristics including  
the maximum transverse velocity amplitude and the phase speed are measured as a function of distance along the structure's length. Solar magneto-seismology is applied to these measured parameters 
to obtain diagnostic information on key plasma parameters (e.g., magnetic field, density, temperature, flow speed) 
of these localised waveguides. The magnetic field strength of the mottle along the $\sim$2 Mm length is found to decrease by a factor of 12,     
while the local plasma density scale height is $\sim$280$\pm$80~km.

\end{abstract}

\keywords{Waves --- magnetohydrodynamics (MHD) --- Magnetic fields --- Sun: Atmosphere --- Sun:  chromosphere --- Sun: oscillations}

\section{Introduction}

Chromospheric  fine-scale structures such as limb spicules, 
on-disk mottles and dynamic fibrils are 
among the most popular objects for study
in solar physics today. 
These jet-like plasma features, formed near the network boundaries, 
can protrude  into the transition region and low corona  \citep{beck1,beck2,ster, dep06} and act as conduits for 
channeling energy and mass from the solar photosphere into the upper solar atmosphere and the solar wind \citep{dep,dep06,mort11}.

Recent ground-based and space-borne observations have shown a plethora of waves and 
oscillations in these structures \citep{kukh, zaq2, dep1, he, he2, zaq, okamo, kur1, mort1, math13}. 
These oscillations are usually observed as periodic transverse displacements    
\citep{zaq, okamo, piet, kur1, mort1}. The observations support the idea that the chromospheric fine-structures 
can be modelled  as thin, overdense magnetic flux tubes
that are waveguides for the transverse oscillations 
with periods which have an observational upper bound limited by their finite visible lifetime.
This is also supported by 3-D numerical modelling of the chromosphere \citep{len12}.  
In this regard, the observed transverse oscillations  have been interpreted  as 
fast kink MHD waves \citep{spru,erdfed}.

Despite a number of recent advances in the field, a detailed study of the properties and physical nature  of the chromospheric fine-structures 
remains a challenging observational task. The propagating or standing nature of the oscillations  
are crucial in understanding the role of these waves in the energy balance of the solar 
atmosphere and their contribution to the atmospheric heating process.
Both propagating (upward and downward) and standing transverse 
oscillations have been reported  in limb spicules \citep{zaq2, zaq, he, he2, okamo}. 
In addition, a few instances of propagating oscillations were recorded as well in chromospheric
mottles   \citep{kur1, mort1}, 
that are believed to be the disk counterparts of limb spicules \citep{tsirop97, zach2, hanst3, scul09, voort09}.

The study of MHD wave properties in spicular structures opens new dimension to chromospheric plasma diagnostics using the tools of solar magnetoseismology (SMS), a field that has recently emerged 
\citep{erd1, and09, rud09, tar09}.
One of the techniques developed estimates the variation of magnetic field strength and plasma density along the chromospheric magnetic flux tube 
from the properties of kink waves 
\citep{verth1}. 
 This approach has been successfully applied to Hinode/SOT Ca {\sc{ii}} H limb spicule observations \citep{verth1}.

In this paper, we present results on transverse oscillations observed in the H$\alpha$  on-disk and quiet-solar chromospheric mottles. 
Our works provide evidence for upward and downward propagating and standing waves.  
In the  case of propagating sample wave characteristics such as maximum transverse velocity 
amplitude and phase speed are measured as a function of distance along the structures length.
The wave properties are used to estimate plasma parameters along the waveguide by employing the SMS approach.

\begin{figure*}[t]
\begin{center}
\includegraphics[width=15.4cm]{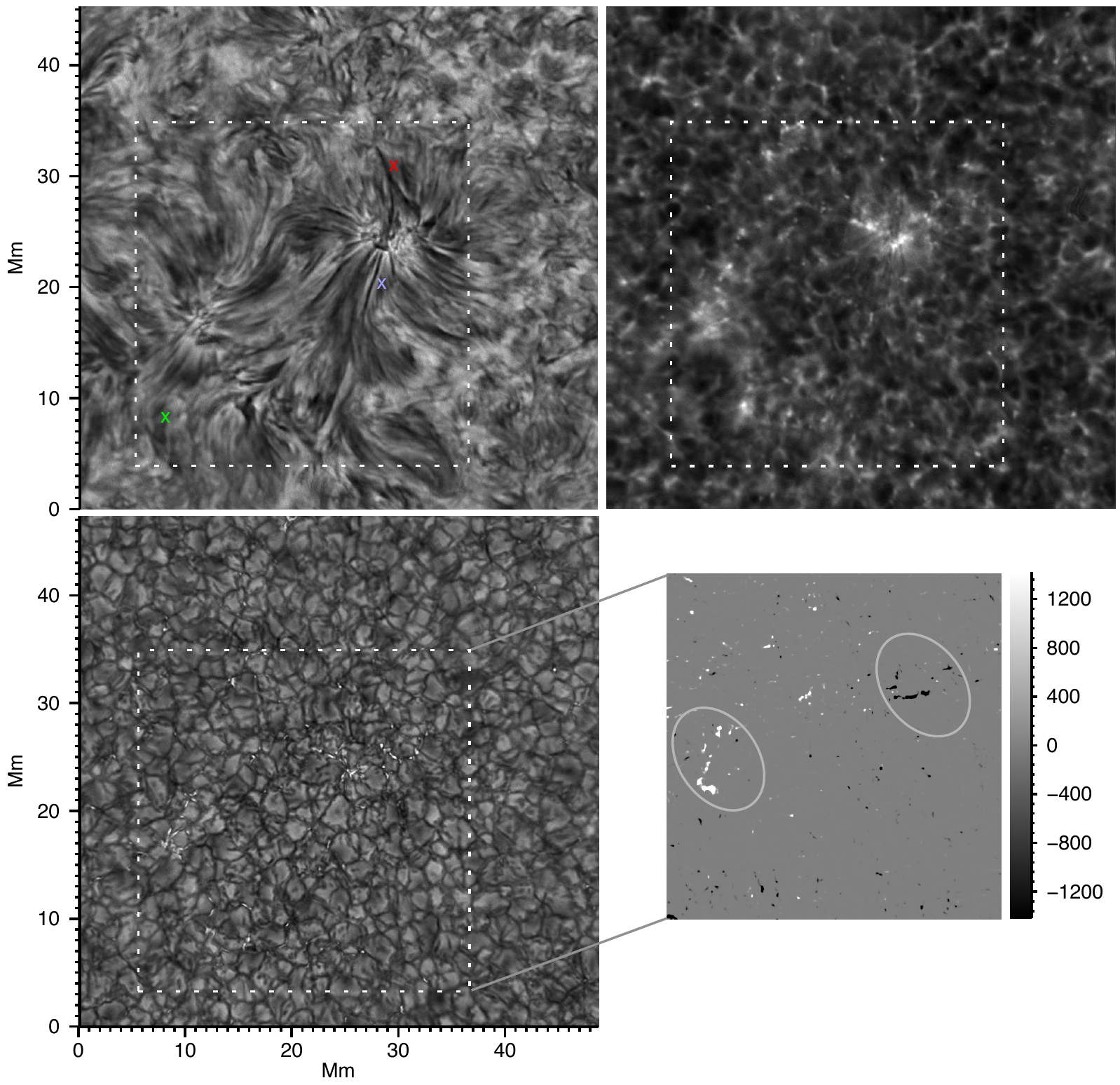}
\end{center}
\caption{Co-spatial and co-temporal ROSA images in H$\alpha$ core (top left), Ca {\sc{II}} K core (top right), G band (bottom left) together with a LOS  magnetogram (bottom right). 
The rosettes, where the mottles are concentrated, are highlighted with white boxes.  Crosses in the H$\alpha$ image indicate the positions of the mottles selected for the analysis.  
The ellipses in the 
LOS magnetogram denote patches of magnetic flux concentrations, with the color scale indicating the magnetic field strength in Gauss. 
The H$\alpha$ image shows that the 
footpoints of the chromospheric fine structures correspond to Ca {\sc{ii}} K and G band bright points and strong magnetic flux concentrations 
that highlight the boundaries of the supergranular cell.}  
\label{fig1}
\end{figure*}

\section{Observations and data reduction}
 
Observations were undertaken between  13:46 - 14:40~UT on
28 May 2009 at disk centre with the Rapid Oscillations in the Solar Atmosphere 
\citep{jess10a}  imaging system, 
and with the Interferometric Bidimensional Spectrometer \citep{cav06},
mounted at the Dunn Solar Telescope (DST) at the National Solar 
Observatory, New Mexico, USA. The ROSA dataset includes simultaneous imaging in the H$\alpha$ core at 6562.8~{\AA} (bandpass 0.25~{\AA}), 
Ca {\sc{II}} K core at 3933.7~{\AA} (bandpass 1.0~{\AA}), G band at 4305.5~{\AA}, bandpass (9.20~{\AA}), and line-of-sight (LOS) magnetograms. 
High-order adaptive optics were applied throughout 
the observations \citep{rim}. The images were reconstructed by implementing the speckle algorithms of  \citet{wog} followed by de-stretching. 
These algorithms have removed the effects of atmospheric distortion from the data. The effective cadence after 
reconstruction is reduced to 4.2243~s for H$\alpha$ and Ca {\sc{ii}} K. 
Observations were obtained with a spatial sampling  of $0.069''$/pixel corresponding to a spatial resolution of  $0.21''$ over the $62''\times62''$ field-of-view (FOV). 

LOS magnetograms were constructed  using the left- and right-hand circularly polarised light obtained 125 m{\AA} into the blue 
wing of the magnetically-sensitive Fe  {\sc{i}}  line at 6302.5 {\AA}. 
A blue-wing offset was required to minimise granulation contrast, while conversion of the 
filtergram into units of Gauss was performed using simultaneous SOHO/MDI magnetograms (see discussion in Jess et al.~2010b).

IBIS undertook simultaneous Na {\sc{i}} D$_1$ core imaging at  5895.94~{\AA} 
with a spatial sampling of  
$0.083''$/pixel over the same FOV. The IBIS data have a post-reconstruction cadence of 39.7~s.
Despite difficulties in interpreting the Na {\sc{i}} D$_1$ line formation height, 
it is suggested that it is formed in the upper photosphere/lower chromosphere \citep{eibe01,fin04}.
Doppler wavelength shifts of the Na {\sc{i}} D$_1$ line profile minimum were used to construct LOS velocity maps of the same FOV  \citep{jess10b}.

\section{Results and analysis}

\begin{figure*}[t]
\begin{center}
\includegraphics[width=15.31cm]{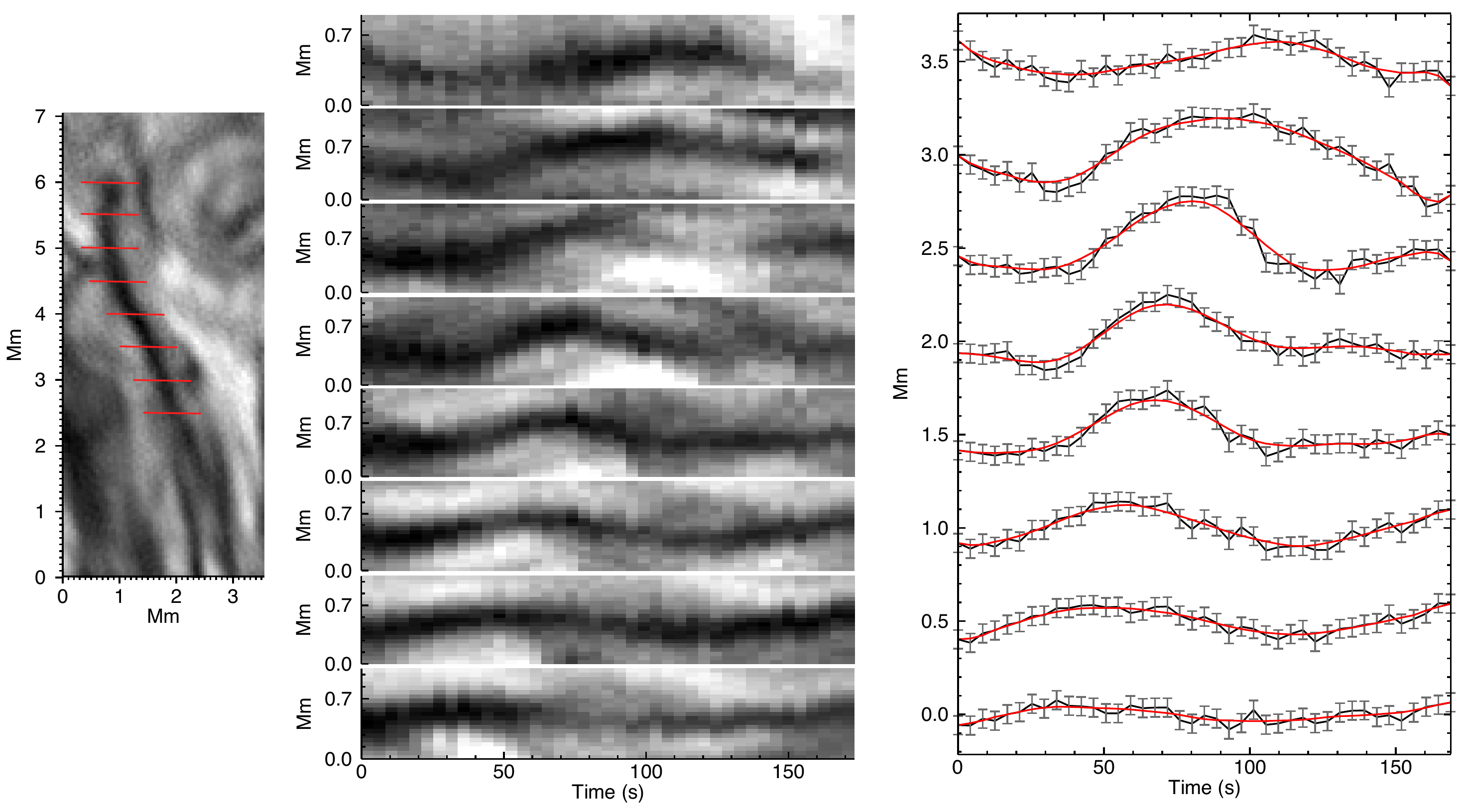}
\end{center}
\caption{Left panel: An expanded view of the mottle from its bottom to top in the image plane located near the red cross in Figure~\ref{fig1}. Red horizontal lines, separated by $\sim$0.5 
Mm, indicate the locations chosen to analyse transverse displacement.
Middle panel: Time-distance diagrams generated from the eight locations indicated by the red lines in the image on the left. 
Right panel: The displacement time series with associated errors. The black lines are the centroid of a Gaussian fit to the cross-sectional flux profiles
of the mottle at each time frame of the transverse cross-cuts smoothed by a $\sim$25 km width (red lines).} 
\label{fig2}
\end{figure*}
\begin{figure*}[t]
\begin{center}
\includegraphics[width=15.31cm]{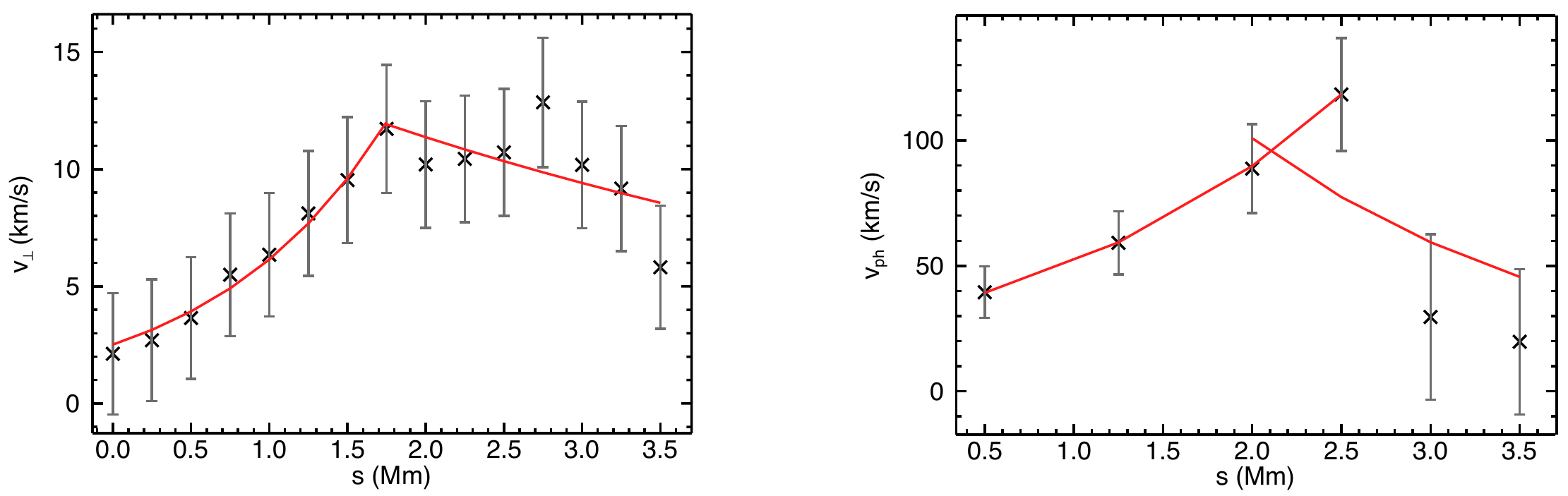}
\end{center}
\caption{Maximum transverse velocity amplitude (left panel) and phase speed (right panel) as a function of distance ({\it s}) along the mottle. The results are  fitted with exponential functions (red lines).
Errors in the transverse velocity and phase speed measurements are also shown.} 
\label{fig3}
\end{figure*}

Figure~\ref{fig1} shows co-spatial and co-temporal snapshots in the H$\alpha$ core and in Ca {\sc{ii}} K, and G band, where the FOV   
covers a quiet Sun region near disk centre. The H$\alpha$ image contains a large rosette structure located near the centre  
(see the top left panel of Figure~\ref{fig1}).
Rosettes are clusters of elongated, dark H$\alpha$ mottles 
expanding radially around a common centre over internetwork regions \citep{zach2,ts1,roup1}. 
An additional three smaller rosettes are visible in the lower left of the boxed area (the top left panel of Figure~\ref{fig1}). 
The roots of the rosettes are  co-spatial with Ca {\sc{ii}} K  brightenings, G band bright points and strong magnetic field 
concentrations which outline the boundaries of supergranular cell highlighted with the dashed box in Figure~\ref{fig1}.
A LOS magnetogram of the FOV shows that the supergranular cell  boundary consist of 
opposite polarity magnetic field concentrations (bottom right panel of Figure~\ref{fig1}). 

The application of time-distance analysis to the H$\alpha$ images reveals that the mottles display 
transverse motions perpendicular to their axis, usually interpreted as transverse MHD kink motions  \citep{spru,edrob, erdfed}.
Periodic transverse displacements of three different mottles, marked with  crosses in Figure~\ref{fig1}, have been selected for further analysis.  

Figure~\ref{fig2} shows a more detailed view of the mottle located near the  red cross in Figure~\ref{fig1}. 
The projected length of the structure is $\sim$ 4 Mm, 
with a resolved average width of ~350 km
and lifetime of $\sim$3 min.  
Red lines across the structure indicate the locations of the cross-cuts 
used to study the transverse oscillations.
We use the cross-cuts to generate two-dimensional time-distance  
(t-d) diagrams that reveal the transverse motions at each point along the mottle's length. These t-d samples are plotted in the middle panel of Figure~\ref{fig2}. 
The shift detected in the signal, as a function of time, indicates that the observed oscillation is due to a propagating wave. 
Displacements are determined 
by fitting a Gaussian function to the cross-sectional flux profile for each time frame of the transverse cross-cuts (right panel of Figure~\ref{fig2}).
This method can determine the position of the structure's centroid to within one pixel and thus has an error of $\pm$50~km.
It should be noted that  in Figure~\ref{fig2} there are only eight cross-cuts (separated by 0.5 Mm), the corresponding 
time-distance diagrams and time series, respectively. 
We note that we generated and analysed the t-d diagrams for 15 cross-cuts separated by 0.25 Mm but we chose to show here
only 8 of those for presentation purposes.
A linear trend was subtracted from the displacement time-series to obtain the periodic motions. 
The time series was fitted with a harmonic function at each position along the mottle 
from which the periods of the wave are derived with a median value  of  $P\approx\mathrm{120\pm10~s}$.

We measured the maximum transverse displacements 
at each of the 15 positions along the $\mathrm{\sim3.7~Mm}$ mottle length  (left panel of Figure~\ref{fig2}).  
The maximum transverse velocity amplitudes are derived using $v_{\bot}=2\pi \xi_{\bot}/P$,
where $\xi_{\bot}$ and $P$ are 
the maximum transverse displacement and period of the oscillation, respectively.
Uncertainties in $v_{\bot}$ are estimated from the error in $\xi_{\bot}$ and the  
standard deviation of $P$. 
The results with the associated error of each data point are plotted in the left panel of  Figure~\ref{fig3}. 
This figure clearly shows that the velocity amplitude is increasing up to around $\mathrm{2~Mm}$ (Figure~\ref{fig3}) and then decreases. Due to the different trends  
the first 8  and last 7 data points are fitted separately by an exponential function 
of the form  $v_{\bot}(s)=v_{\bot,0}\exp(s/A_{2})$, where  ${\it s}$  
is the distance along the mottle. 
We note that the exponential 
function has the best-fit (with 95 $\%$ confidence level) for the first 8 datapoints.
We obtain   $v_{\bot,0}=2.5\pm1.0~\mathrm{km~s^{-1}}$  and  $A_{2}=1.12\pm0.35~\mathrm{Mm}$ 
for ${\it s}\leqslant1.75~\mathrm{Mm}$ and   $A_{2}=-5.3\pm4~\mathrm{Mm}$  for ${\it s}>1.75~\mathrm{Mm}$.
Errors in these fitting parameters are their 1$\sigma$ uncertainties derived from the fitting algorithm, which use the measurement error for each data point.

The phase speed of the transverse motions can be evaluated from the 
time-delay in the signals obtained at different positions along the mottle.  
From the displacement time series  we can measure the time coordinates of the maximum transverse displacements. 
A phase speed between two positions along the structure can be calculated as  $v_{ph}=L/t_l$,  
where $L$ is the distance between two selected heights and   $t_{l}$ is the time delay between the location of the maximum displacements.    
The time at the location of the maximum transverse displacement in the time series can be estimated within one temporal resolution element of 4.224~s,  
and the maximum phase speed which can be resolved for the length $L$ along the structure is  $v_{ph_{max}}\sim L/4.224$.   
We determine a reliable  phase speed of the transverse wave for six consecutive segments along the mottle length  
and results (with corresponding measurement errors) are plotted in the right panel of  Figure~\ref{fig3}.  
The phase speed is $\mathrm{\sim40~km~s^{-1}}$ near the lower part of the mottle and increases to $\sim\mathrm{120~km~s^{-1}}$ at 2.5~Mm,  
then it decreases again towards the end of the structure (right panel of Figure~\ref{fig3}).  
We fit the data points using an exponential function 
of the form  $v_{ph}(s)=v_{ph,0}\exp(-s/A_{1})$  
with $v_{ph,0}=29.7\pm6.7~\mathrm{km~s^{-1}}$,  $A_{1}=-1.81\pm0.24~\mathrm{Mm}$  for ${\it s}\leqslant2.5~\mathrm{Mm}$ and   $A_{2}=1.88\pm0.3~\mathrm{Mm}$  for ${\it s}>2.5~\mathrm{Mm}$.


\begin{figure*}[t]
\begin{center}
\includegraphics[width=16.0cm]{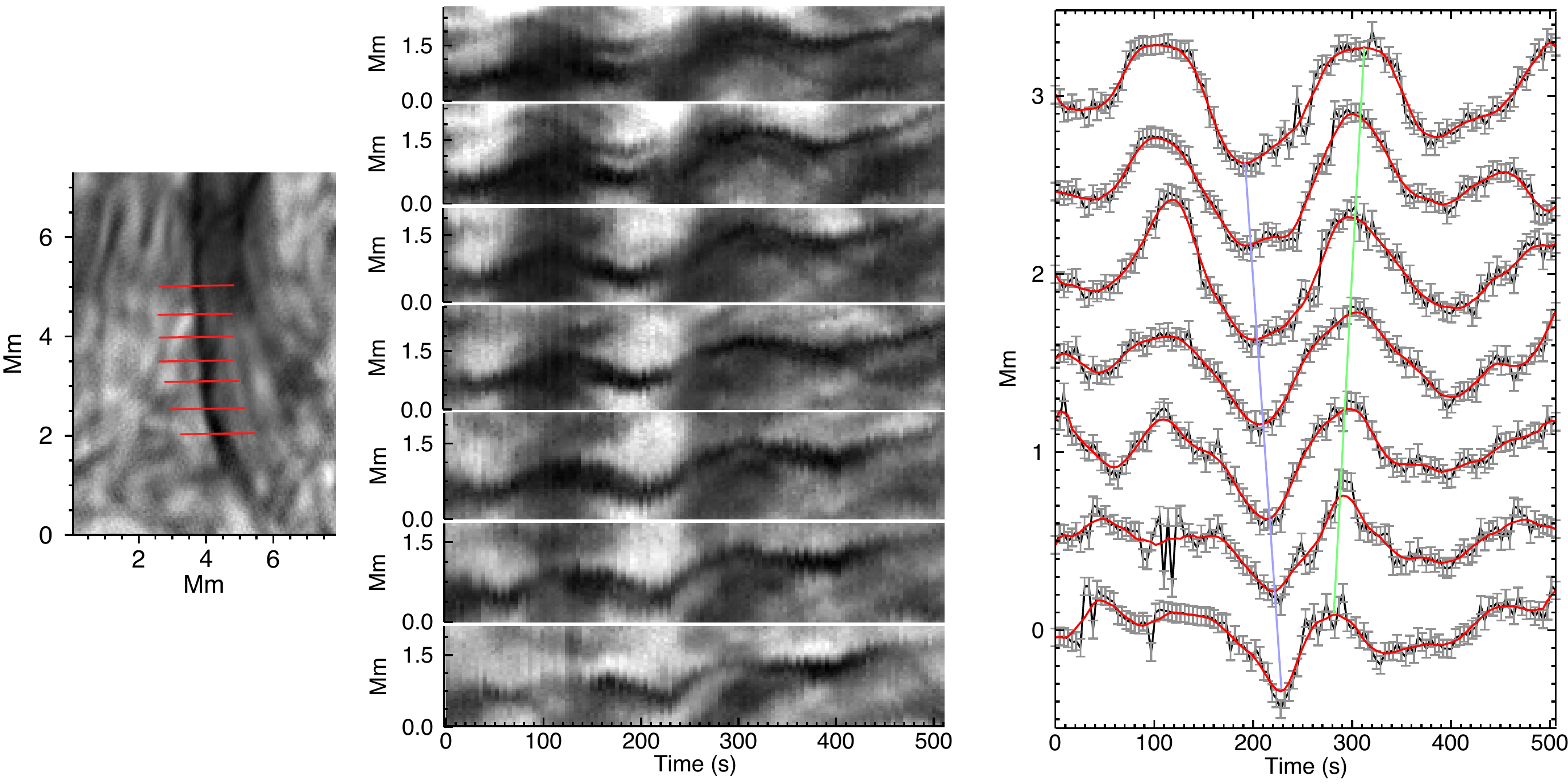}
\end{center}
\caption{Left panel: A detailed view of the mottle from its bottom to top in the image plane located near the green cross  in Figure~\ref{fig1}. Red short lines separated by $\sim$0.5 
Mm  indicate the locations on the mottles where time-distance plots depicted on the middle panel (from bottom to top) are generated.
Right panel: The displacement time series with error bars. Diagonal lines highlight the downward (blue) and upward (green) propagating motions.}
\label{fig4}
\end{figure*}

\begin{figure}[t]
\begin{center}
\includegraphics[width=8.6 cm]{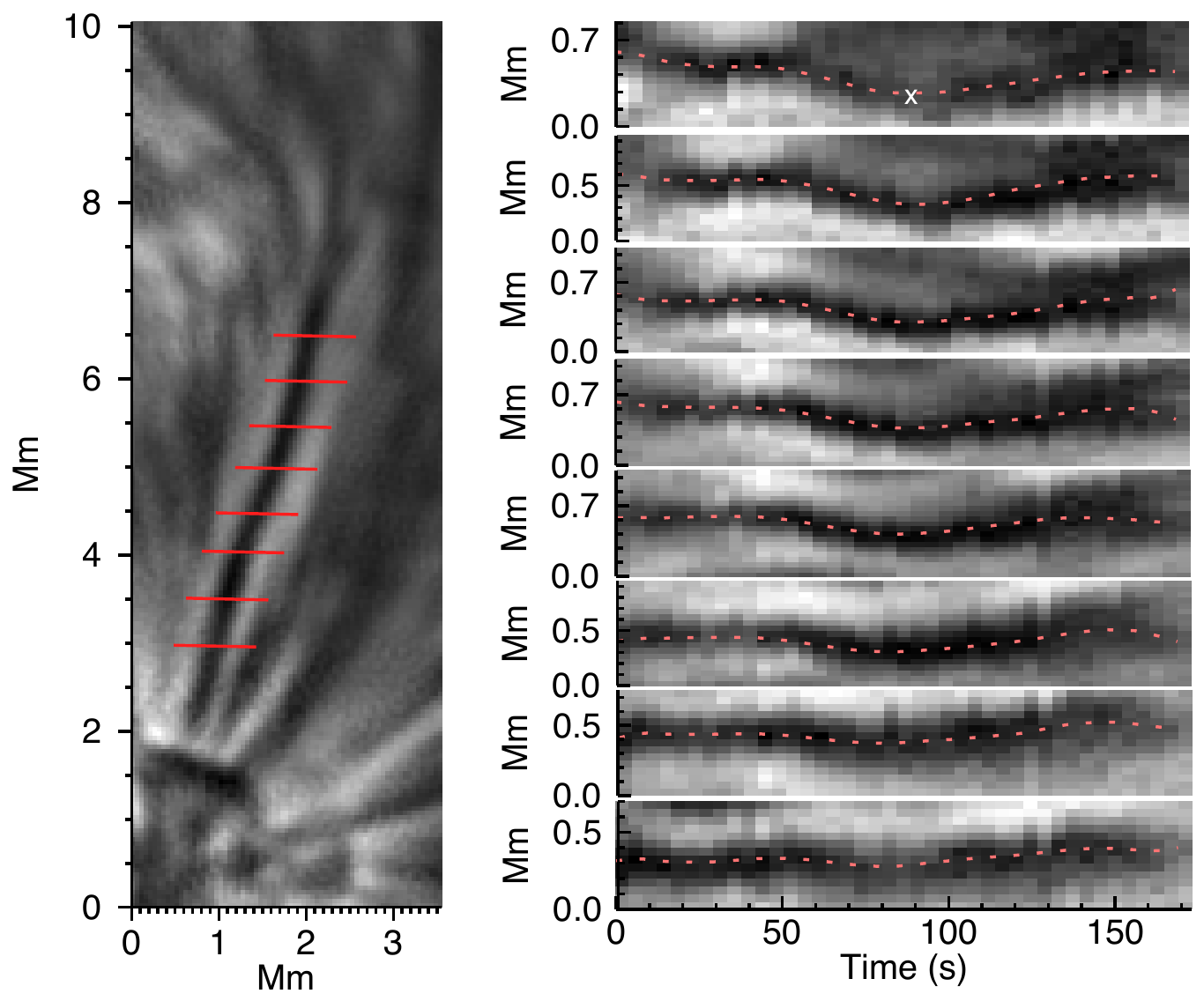}
\end{center}
\caption{The mottle from its bottom to top in the image plane located near the blue cross  in Figure~\ref{fig1}. Cross-cuts, separated by $\sim$0.5 
Mm,  indicate the locations on the mottle where the displacement is further analysed (left panel).
The pink dotted lines are the centroids of a Gaussian fit to the cross-sectional flux profiles of the mottle at each time step 
(right panel).} 
\label{fig5}
\end{figure}

\begin{figure*}[t]
\begin{center}
\includegraphics[width=17.9cm]{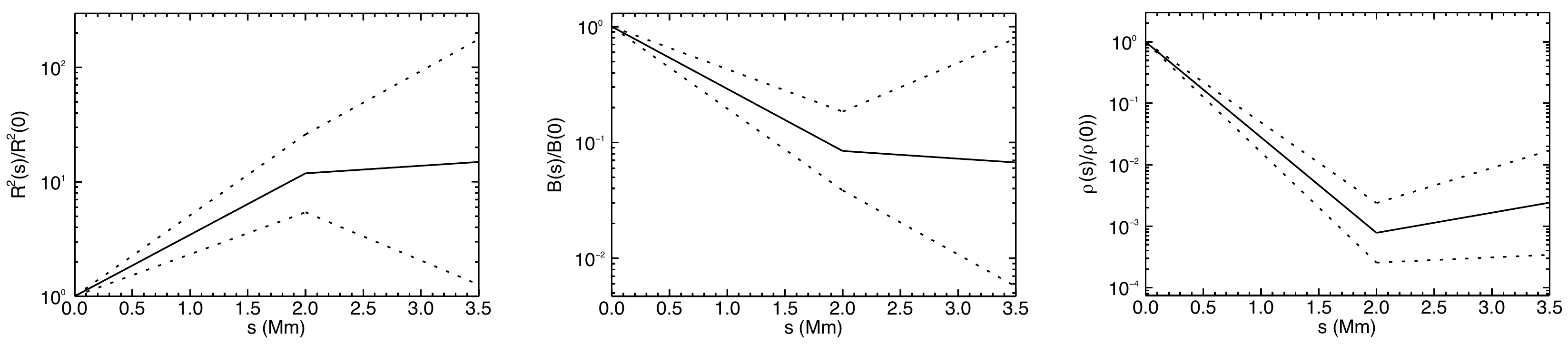}
\end{center}
\caption{Normalised area expansion (left), magnetic field strength (middle) and plasma density (right),
estimated using the techniques of magnetoseismology, plotted as a function of length along the waveguide shown in Figure~\ref{fig2}.
The dotted lines indicate the region of uncertainty due to the $1\sigma$ error of $A_1$ and $A_2$.}
\label{fig6}
\end{figure*}

\begin{figure}[b]
\begin{center}
\includegraphics[width=8.2 cm]{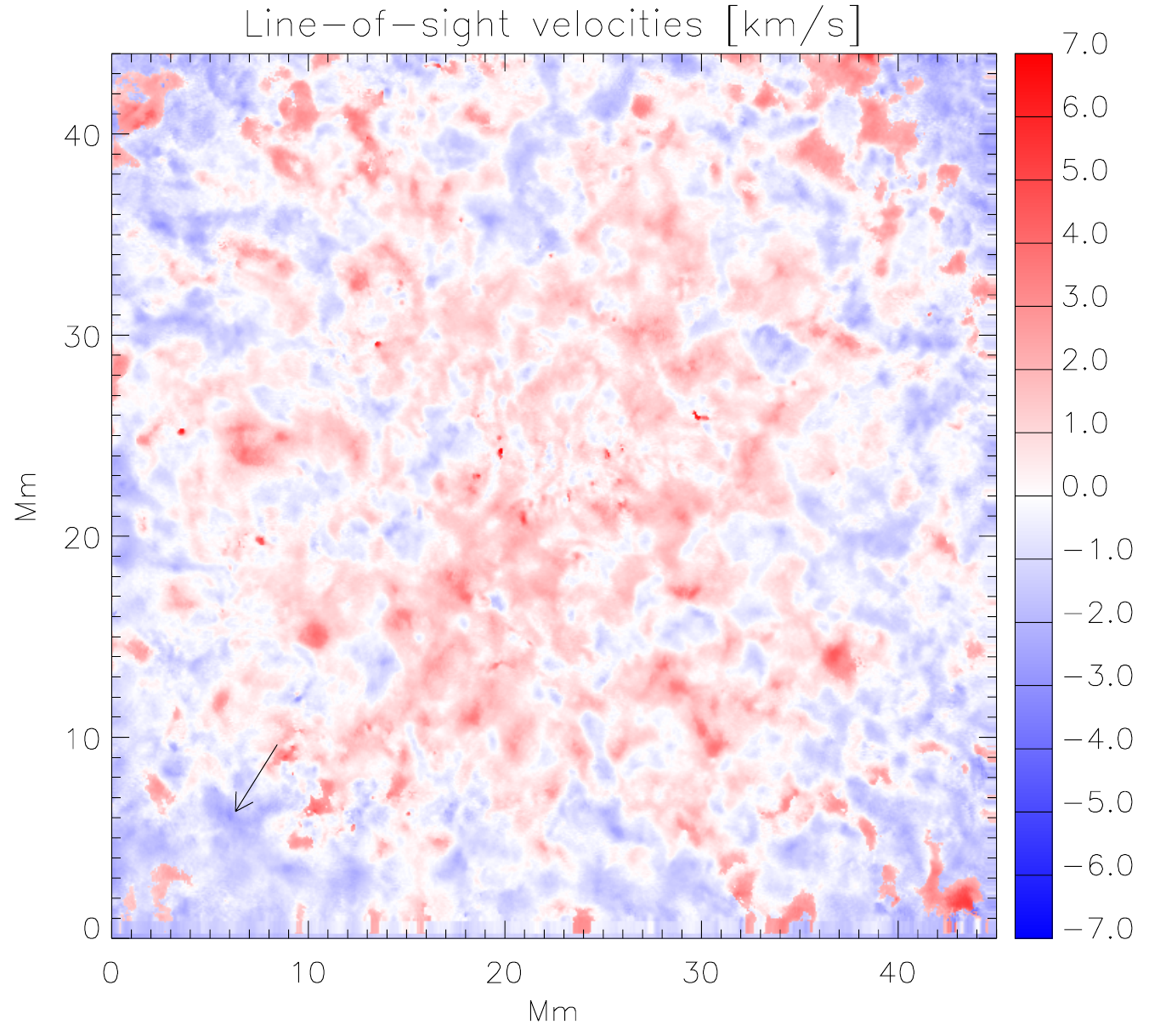}
\end{center}
\caption{Snapshot of the LOS velocity map of the full FOV. 
Red and blue colors represent positive doppler velocities (downflows)  and negative doppler velocities (upflows), respectively.
The arrow indicates the position of the mottle analysed in Figure~\ref{fig4}.}
\label{fig7}
\end{figure}

Figure~\ref{fig4} shows the H$\alpha$ mottle (left panel) located near the  green  cross of Figure~\ref{fig1} 
and its transverse displacements at different positions along its length (middle and right panels of Figure~\ref{fig4}). 
Time series, obtained using a similar method to that described for the first mottle in Figure~\ref{fig2},
highlight the  upward  and  downward  propagating waves with a period of $P\mathrm{\approx180\pm10~s}$ 
(see the green and blue diagonal  lines in the right panel of Figure~\ref{fig4}). 
The average  phase speed and maximum transverse velocity along the length of the mottle are 
$\mathrm{101\pm14~km~s^{-1}}$, $\mathrm{8.8\pm3.1~km~s^{-1}}$, $\mathrm{79\pm8~km~s^{-1}}$, 
and $\mathrm{11.4\pm3.3~km~s^{-1}}$  for the upward and downward propagating waves, respectively.
Unfortunately, for this example large uncertainties in the transverse velocity and phase speed measurements 
do not allow us to study their variation as a function of distance.

Figure~\ref{fig5}  shows the chromospheric structure  marked with a blue cross in Figure~\ref{fig1} and its transverse displacement with a period of   $P\approx131\pm\mathrm{15~s}$. 
We detect a marginal delay, of about 10~s, in the oscillation signals at the lower and upper positions
(Figure~\ref{fig5}). This time delay, combined with a distance of around 3.5~Mm  between these positions (see left panel of Figure~\ref{fig4}), 
gives a propagating speed of more than 350~km/s, 
too high a value for what might be expected for the phase speed of the kink waves in chromospheric mottles. 
We believe that this high speed may be caused by the standing wave pattern generated by the superposition of two oppositely directed waves.

\section{Magnetoseismological diagnostics} 

A novel solar magnetoseimology tool that allows us to determine the variation of 
magnetic field and plasma density along a chromospheric structure using the characteristics of 
kink oscillations has been developed by \cite{verth1}. Based on their approach,  
if the kink speed $v_{ph}(s)$, (which is the phase speed of the kink wave) and maximum transverse velocity amplitude $v_{\bot}(s)$ are 
estimated from observations, the expansion rate of the magnetic flux tubes can be 
derived from the solution of the kink wave governing equation (see Equation 1 of Verth at~al. 2011). 
The flux tube radius as a function of $s$ is given by  
\begin{equation}
R(s) =R(0)\exp{\left(s\over H\right)}~,
\label{eq1}
\end{equation}
where $H=2A_1A_2/(2A_1+A_2)$ and $A_1$, $A_2$ are fitting parameters defined from the measured 
$v_{ph}(s)$ and $v_{\bot}(s)$, and $R(0)$ is the flux tube radius at the lowest position  \citep{verth1}. 
On the other hand, from magnetic flux conservation  $B(s)\propto1/R^2(s)$, where  $B(s)$ 
is the average magnetic field strength, and the  variation of magnetic field along the 
flux tube can be estimated. Furthermore, from the kink speed and magnetic field variations, the plasma density along 
the structure can also be determined using  
\begin{equation}
{\rho(s)}\propto {B^2(s)\over v_{ph}^2(s)}, 
\label{eq2}
\end{equation} 
where $\rho(s)$ is the average of the internal and external plasma densities (see Verth et~al. 2011 for further details). 

Inspired by this work, we estimate these parameters for the on-disk mottle presented on Figure~\ref{fig2} 
using the same SMS tool. We employ the functions of $v_{ph}(s)$ and $v_{\bot}(s)$ 
found in the previous section using the exponential fit for the measured phase travel time and velocity amplitudes.  
From these functions, which define $A_1$ and $A_2$, and from Equation~(1) 
we make piecewise estimate the normalised area expansion of the flux tube. 
The magnetic field variation can also be evaluated from the  area expansion and magnetic flux conservation law 
(middle panel of Figure~\ref{fig6}).  Furthermore, the normalised plasma density  along this mottle, estimated from Equation~(2),  
is shown in the right panel of Figure~\ref{fig6}.  
\\

\section{Discussion and conclusions}

Several observational and theoretical studies suggest that the transverse MHD kink waves observed in 
chromospheric structures can be excited by granular buffeting, 
global oscillations,  mode conversion or torsional motions in the photospheric bright points 
where chromospheric fine structures are anchored \citep{rob1,spru1,hol1,has1,dep,jess2, mort3}. 
By tracking the oscillation signals at different positions, 
we detected  both propagating and standing wave modes along chromospheric mottles
which appear to be rooted in regions with strong magnetic field concentrations  
(Figure~\ref{fig1}).  Upward propagating waves  with periods of $\mathrm{\sim120~s}$  
are detected in one of the mottles (Figure~\ref{fig2}).  The analysis shows that the phase speed and transverse velocity amplitude 
rise exponentially with distance along the mottle length up to about $\mathrm{2.5~Mm}$ before they begin to decrease (Figure~\ref{fig3}).  
MHD wave theory suggests that the variation of these wave characteristics are controlled 
by changes in plasma parameters such as density and magnetic field strength along the waveguide. 
A decrease in the plasma-$\beta$ will result in the mottle plasma parameters gradually becoming 
dominated by the magnetic field, causing the observed growth of the phase speed (Figure~\ref{fig3}). 
At a height of around $\mathrm{2-2.5~Mm}$, the magnetic canopy is formed \citep{solan,wede,tsiro3}.  
This is the layer  where the gas and magnetic pressures are equal  
($\beta\approx1$) and where the mode conversion through e.g. non-linear interactions 
can occur   \citep{ros1, bog1, has2, schu, kur2}. The amplitude of the transverse motions 
increases and  at about $\mathrm{2~Mm}$ reaches $\mathrm{250~km}$ (Figure~\ref{fig2}), 
a value similar to the waveguide width of 350~km. 
Hence,  the observed fast kink wave mode may become  nonlinear  near the canopy area, 
which may lead to mode conversion, and thus energy transfer between nonlinear kink modes and longitudinal waves.  
This can result in the observed decrease of phase speed and transverse velocity at a higher length along the structure (Figure~\ref{fig3}).

Waves that propagate from the lower chromosphere 
into the transition region may undergo 
reflection at the top of the canopy due to the sharp density gradient  \citep{hol2, ros1, kur2, fuji}.  
The conditions for the reflection of the kink waves are defined by the local cut-off period, the highest period that is allowed to propagate.        
Following the Kneser oscillation theorem \citep{knes}, if the phase speed is  
increasing with height, the governing kink wave equation processes a cut-off  
which can be calculated as $P_c\approx4\pi \Delta s/\Delta v_{ph}$, 
where $\Delta v_{ph}$ is the change of the phase speed as a function of distance $s$, and   $\Delta s$ is the distance between two selected points.  
For the waveguide presented in Figure~\ref{fig2} this corresponds to a cut-off period of $\mathrm{\sim314~s}$. 
This value is much higher than the observed kink wave period ($\mathrm{\sim120~s}$) suggesting that 
the observed wave, and waves with periods less then the estimated cut-off,  
should propagate into the upper chromospheric layers  
without reflection. The high cut-off period indicates that the chromospheric mottles  
could allow the propagation of long period ($\mathrm{P>3~min}$) 
transversal (kink and Alfv\'en) waves as well. 
Those waves are observed in the corona and are thought to be an important contributor to the coronal heating, at least in the case of the quiet Sun \citep{mac12}.  


It appears that the upward and downward directed waves (green and blue diagonal  lines in the right panel of Figure~\ref{fig4})  
have an approximately constant phase speed ($\Delta v_{ph}\approx0$) along the $\mathrm{\sim3~Mm}$  mottle length.
Almost constant phase speeds for torsional waves were also detected along limb spicules  \citep{dep12,sek13}.
A constant phase speed suggests that there is no cut-off period, 
i.e. waves of any period can propagate along those 
fine-scale structures in the chromosphere. However, the downward propagating wave detected along the mottle presented in Figure~\ref{fig4}   
may be formed as a result of reflection of the upward propagating wave  
at the transition region boundary  or in the corona. 
Furthermore, for this mottle we measured $v_{up}\approx101\mathrm{~km~s^{-1}}$ and $v_{down}\approx79\mathrm{~km~s^{-1}}$, 
where   $v_{up}$ and  $v_{down}$ are the phase speeds of upward and downward propagating waves, respectively. 
This difference could be a result of plasma flow along the mottle.  
In the presence of flow the upward and downward kink speeds are modified by the flow as follows,
\begin{equation}
v_{up}=v_{ph}+U, 
\label{eq3}
\end{equation} 
\begin{equation}
v_{down}=v_{ph}-U, 
\label{eq4}
\end{equation} 
{where, $v_{ph}$ is the kink speed for the mottle with no flow, and $U$ is the flow speed. 
This suggests a plasma flow along this mottle of $U\approx 11\mathrm{~km~s^{-1}}$ in the upward direction.
A snapshot of the LOS velocity map of the studied region, obtained from Doppler wavelength shifts of the $\mathrm{Na~I~D_1}$ profile,  
shows that the lower chromosphere is dominated by the flow patterns (Figure~\ref{fig7}).  
Values of the flow speed vary from around $-4.2\mathrm{~km~s^{-1}}$ (upward) to $7\mathrm{~km~s^{-1}}$ (downward) with an error of $\pm0.5~\mathrm{km~s^{-1}}$ (Jess et al.~2010b). 
The average upflow  LOS velocity near the footpoint  of this mottle  
is around  $\mathrm{-4~km~s^{-1}}$ (Figure~\ref{fig7}). This value is consistent with the seismologically estimated phase speed, 
$\mathrm{\sim10~km~s^{-1}}$, which is the horizontal component and thus may be higher depending on the structure's inclination. 
Recently, \cite{vis12} measured upflow/downflow velocities  within the range of  $\mathrm{-5.7~to~+13.5~km~s^{-1}}$ 
in chromospheric fibrils, consistent with our observations and SMS estimations.

A superposition of the opposite directed kink waves  may result into a wave with a very high 
phase speed which can be considered as a (partially) standing wave \citep{fuji}.  
The phase speed of the transverse wave shown in Figure~\ref{fig5} is $\mathrm{\sim350~km~s^{-1}}$. 
This value is considerably higher than the local Alfv\'en/sound/kink speeds, indicating that it may 
be the consequence of the superposition of up and down propagating waves.

In Figure~\ref{fig6} we show the normalised estimated area expansion, magnetic field and plasma density variations 
as function of length along the waveguide shown in Figure~\ref{fig2}. 
The area expansion factor along the $\mathrm{\sim2~Mm}$ flux tube length is found to be $\mathrm{\sim12}$ (left panel of Figure~\ref{fig6}), 
with a decrease in the magnetic field strength of the same factor  (middle panel of Figure~\ref{fig6}). 
Unfortunately, even modern high-resolution observations can not yet provide direct, precise  measurements 
of the flux tube expansion rate and magnetic field variation from the photosphere to chromosphere. 
Thus, it is very difficult to compare the results obtained from magnetoseismological techniques with direct measurements. 
We emphasize that spectropolarimetric measurement of chromospheric spicular magnetic field strengths of  
$B_0\mathrm{\sim10-50~G}$ \citep{truj05,cen10},  and the observed footpoint photospheric magnetic field strength  
of $\mathrm{\sim kG}$ (lower right panel of Figure~\ref{fig1}) give comparable factors for the decrease in the field strength.

It must be noted that the H$\alpha$ image of the mottle (left panel of Figure~\ref{fig6}) does not show visual expansion by a factor of 12 (that would correspond to a radius change by a factor of 3.5).
Gaussian fit to the cross-sectional flux profile of the mottle suggests that the width of this structure at around 2~Mm at its length (left panel of Figure~\ref{fig6}) is $\sim$350~km.
Hence, according to our SMS estimation the width at its base is expected to be $\sim$100~km. This is the typical diameter of the G-band bright points 
which are considered as the footpoints of the mottles \citep{croc10}. 
However, 100~km is below the spatial resolution of the ROSA H$\alpha$ filter. 
\cite{def07} has suggested 
that the  solar threadlike structures' expansion may 
not be seen visually (through imaging observations) if the threads have sub-resolution width. 
The chromospheric fine structures analysed in this paper are near the resolution limit which could be the reason of the relatively constant apparent width along their length. 
%
%
%
%
%
%

The normalised plasma density from our magnetoseismological study (plotted in the right panel of Figure~\ref{fig6})
shows that the density along the mottle decreases 
by $10^3$ in a 2~Mm length. 
Despite such significant drop in density, the mottle is still visible in the H$\alpha$ images. 
For a resolved dense flux tube in the chromosphere, the intensity is proportional to density, opacity, geometric depth and the source function. 
The source function gives the contribution that the plasma makes to the intensity due to absorption/emission and cannot be determined directly from observation. 
With respect to unresolved (or near resolution) flux tubes there are added complications. 
DeForest (2007) investigated the effect of geometric expansion on intensity for such structures. 
He found that the effect of sub-resolution flux tube expansion results in an apparent constant flux tube width 
and enhanced brightness with height (see Fig. 4 of DeForest 2007) .This simple geometric effect could be true for the dark, absorption H$\alpha$ mottles as well,
%
and hence it may be the explanation to why the upper part of the mottle is visible in H$\alpha$. 
In addition, we would like to emphasise that the density of H$\alpha$ dark mottles estimated by previous work (see e.g. Tsiropoula \& Schmieder~1997) 
is about $10^{10}$~cm$^{-3}$.  
For this value, our SMS density diagnostic method suggests that near the base the density would be approximately $10^{13}$~cm$^{-3}$, 
which is a realistic value according to different atmospheric models \citep{ver81,fon07}.

Density diagnostics provide an estimate for the local plasma density scale height of $H_{\rho}\mathrm{=(280\pm80)~km}$ along the $\mathrm{2~Mm}$ length,  
lower than some previous seismological estimates of the scale height ($\mathrm{\sim700~km}$) in limb spicules 
 \citep{verth1, mak}. The density scale height could be used to estimate the mottle temperature 
 in the isothermal approximation using  $H_{\rho}\approx\mathrm{[T/(1~MK)]~47~Mm}$ \citep{asch04}.
This yields $T=\mathrm{(5957\pm1702)~K}$ for the mottles presented in Figure~\ref{fig4}.
The earlier work of \cite{gio67} estimates the temperatures of the dark mottles to be $T<\mathrm{10000~K}$.  
Based on some parameters given
by the cloud model \cite{tsirop93} derived values in the range 7100-13000~K.  Later on, \cite{tsirop97} claimed that dark mottles 
for the microturbulent velocity around $10\mathrm{~km~s^{-1}}$  
have $T=14000~\mathrm{K}$ 
with standard deviation~$\approx9200~\mathrm{K}$  (see Table 1 of Tsiropoula  \& Schmieder~1997). 
The SMS temperature diagnostic suggests that the particular 
dark mottle analysed here is at the lower end of previous temperature range estimates. 
Providing new insight, SMS suggests the dark mottle is reasonably isothermal along its structure, 
at least up to 2~Mm from its footpoint. However, more SMS dark mottle case studies will be required 
to actually understand how representative the present example is.

The application of SMS diagnostics to the mottle for lengths greater than $\sim$2~Mm, show a decrease in the plasma density and magnetic field gradients. 
Although these features were also found for an off-limb spicule \citep{verth1}, 
we point out that our estimates above 2~Mm carry large uncertainties. 
The observed changes could be caused by the effects of the magnetic canopy. 
The merging flux tubes higher in the atmosphere could alter the rate at which the magnetic field decreases. 
At the canopy level the flux tubes become more horizontal which can change the density stratification along the structure.
We note that the SMS estimates  presented here are more applicable to the local plasma parameters of a particular small-scale flux tube,
and they may not necessarily be considered as typical of  all chromospheric structures.  
However, it has been demonstrated that by studying the variation of phase
speed and transverse velocity of kink waves along 
mottles and fibrils we can understand more completely the dominant plasma properties of chromospheric waveguides.
Furthermore, a wealth of statistics for phase speed variations can provide typical values for the cut-off period. 
Transverse oscillations which are ubiquitous in the chromosphere  \citep{kur1, mort1} are likely to be separated in propagating and non-propagating waves by the cut-off period.
Hence, it could be crucial to estimate how much kink wave energy is transported into the corona and what is trapped in the chromosphere.

\begin{acknowledgements}
This work is supported by the UK Science 
and Technology Facilities Council (STFC).
RE acknowledges
M. K\'eray for patient encouragement and is also grateful to NSF Hungary (OTKA K83133).
We thank the Air Force Office of Scientific Research, Air Force  Material Command, USAF for sponsorship under grant number FA8655-09-13085.
G.V. acknowledge the support of the Leverhulme Trust. 
\end{acknowledgements}

\end{document}